\documentclass[prb,reprint,amsmath,amssymb,showpacs,superscriptaddress,floatfix,longbibliography]{revtex4-1}
\usepackage{graphicx,epsfig}
\usepackage{amsmath}
\usepackage{xcolor}
\usepackage[breaklinks=true,colorlinks,citecolor=blue,linkcolor=blue,urlcolor=blue]{hyperref}
\usepackage{subfigure}

\def\be{\begin{equation}}
\def\ee{\end{equation}}

\def\ber{\begin{eqnarray}}
\def\eer{\end{eqnarray}}

\begin{document}
\title{Thickness and electric field dependent polarizability and dielectric constant  in phosphorene}

\author{Piyush Kumar}
\affiliation{Dept. of Electrical Engineering, Indian Institute of Technology Kanpur, Kanpur 208016, India}
\author{B. S. Bhadoria}
\affiliation{Dept. of Physics, Bundelkhand University, Jhansi, 284128, India}
\author{Sanjay Kumar}
\affiliation{Dept. of Physics, Indian Institute of Technology Kanpur, Kanpur 208016, India }
\author{Somnath Bhowmick}
\affiliation{Dept. of Materials Science and Engineering, Indian Institute of Technology Kanpur, Kanpur 208016, India}
\author{Yogesh Singh Chauhan}
\affiliation{Dept. of Electrical Engineering, Indian Institute of Technology Kanpur, Kanpur 208016, India}
\author{Amit Agarwal}
\affiliation{Dept. of Physics, Indian Institute of Technology Kanpur, Kanpur 208016, India }
\date{\today}

\begin{abstract}
	{Based on extensive first principle calculations, we explore the thickness dependent effective dielectric constant and slab polarizability of few layer black phosphorene. We find that the dielectric constant in ultra-thin phosphorene is thickness dependent and it can be further tuned by applying an out of plane electric field. The decreasing dielectric constant with reducing number of layers of phosphorene, is a direct consequence of the lower permittivity of the surface layers and the increasing surface to volume ratio. We also show that the slab polarizability depends linearly on the number of layers, implying a nearly constant polarizability per phosphorus atom. Our calculation of the thickness and electric field dependent dielectric properties will be useful for designing and  interpreting transport experiments in gated phosphorene devices, wherever electrostatic effects such as capacitance, charge screening etc. are important. }
\end{abstract}

%
%
\maketitle

\section{Introduction}

Monolayer black phosphorene~\cite{liu2014phosphorene} has emerged as a promising material for p-type FET (field effect transistor) operation, on account of it's direct bandgap of magnitude 1.5 eV and reasonably high hole mobility value, theoretically predicted and experimentally measured to be around 10,000 and 1,000 cm$^2$V$^{-1}$s$^{-1}$, respectively\cite{liu2014phosphorene,li2014black,qiao2014high}. As a consequence, transistors based on phosphorene exhibit excellent I$_{ON}$/I$_{OFF}$ ratio of $10^4-10^5$, as well as carrier mobility comparable to that of MoS$_2$ (200 -- 1,000 cm$^2$V$^{-1}$s$^{-1}$)\cite{liu2014phosphorene,li2014black,xia2014rediscovering}. Individual layers of black phosphorus have puckered honeycomb like structure and they are held together by weak van-der Waals forces, facilitating separation of monolayer or few layers by mechanical or liquid exfoliation\cite{liu2014phosphorene,guo2015,adam2015}. Phosphorene is a highly anisotropic material and values of elastic modulus, electron and hole effective mass, calculated along mutually orthogonal zigzag and armchair direction, have a ratio of 3.5, 6.6 and 42.3, respectively  \cite{qiao2014high}, which is also manifested in experimental observations reporting anisotropic electronic transport and optical properties\cite{liu2014phosphorene,wang2015highly,mao2015, PRB.92.165406}.

It has been shown that the intrinsic bandgap of phosphorene and other layered materials can be modulated by applying external perturbation like strain and electric field \cite{dai2014bilayer,li2014modulation,fei2014strain,ccakir2014tuning,wang2015, Priyank, Barun, Priyank2, Nahas, PhysRevB.91.045433}.  More interestingly, it has also been predicted that  an electric field applied out of the plane of phosphorene can induce a tunable Dirac cone, and additionally induce a normal insulator to topological insulator to metal transition \cite{dolui2015quantum,PTI}. Such studies vividly demonstrate the tunability of electronic properties of multi-layered phosphorene, on application of an external electric field. Motivated by this, in this article we study the thickness and electric field dependent effective dielectric constant of $N$-layers of phosphorene subjected to an out of plane (transverse) electric field. The dielectric constant and polarizability of a material, in addition to being fundamentally  important properties, are also very useful for characterizing the material's  electrostatic properties such as electronic charge screening which determines the strength of Coulomb interaction, capacitance, energy storage capacity, and its transport characteristics \cite{santos2013electric, santos2013electrically, tobik2004electric, yu2008ab, li2015ab}. 

In this article we use density functional theory (DFT) based electronic structure calculations, including van-der Waals interactions, to explicitly show that the dielectric constant of phosphorene, perpendicular to the phosphorene plane,  depends on the number of layers and the strength of the applied transverse electric field. Using these calculations for a few layers (from two to eight) phosphorene, we obtain the bulk dielectric constant and the surface polarizability of phosphorene atoms, which in turn are used to correctly extrapolate the dielectric constant for any number of layers, at least in the low electric field regime. 
The article is organized as follows: in section II we report the structural parameters and electronic properties of multi-layer phosphorene, followed by a discussion of the impact of transverse electric field on charge screening in section III, and it's effect on the slab polarizability and dielectric constant in section IV. Finally we summarize our findings in section V.

\section{Structural and electronic properties of multi-layer phosphorene}
We calculate the structural and electronic properties of multilayered black phosphorene based on DFT calculations, using a plane wave basis set (cutoff energy 100 Ry) and norm conserving Troullier Martins pseudopotential,\cite{troullier1991efficient} as implemented in Quantum Espresso.\cite{giannozzi2009quantum} The electron exchange-correlation is treated within the framework of generalized gradient approximation (GGA), as proposed by Perdrew-Burke-Ernzerhof (PBE).\cite{perdew1996generalized} The dispersion forces among multiple phosphorene layers are taken into account by using non-local van-der Waals exchange correlation functional optB88-vdW.\cite{klimevs2010chemical, dion2004van,thonhauser2007van,roman2009efficient,sabatini2012structural} Brillouin Zone integrations are performed using a $k$-point grid of $16 \times 20 \times 1$. Further we have used  a vacuum region of $20$ \AA~ in the direction perpendicular to the phosphorene plane to prevent any interaction among the spurious replica images. All the structures are fully optimized until the forces on each atom are less than 0.01 eV/\AA. The effect of transverse electric field is simulated using a sawtooth potential, while taking dipole correction into consideration.\cite{bengtsson1999dipole}

\begin{figure}[t]
\includegraphics[width=1.0 \linewidth]{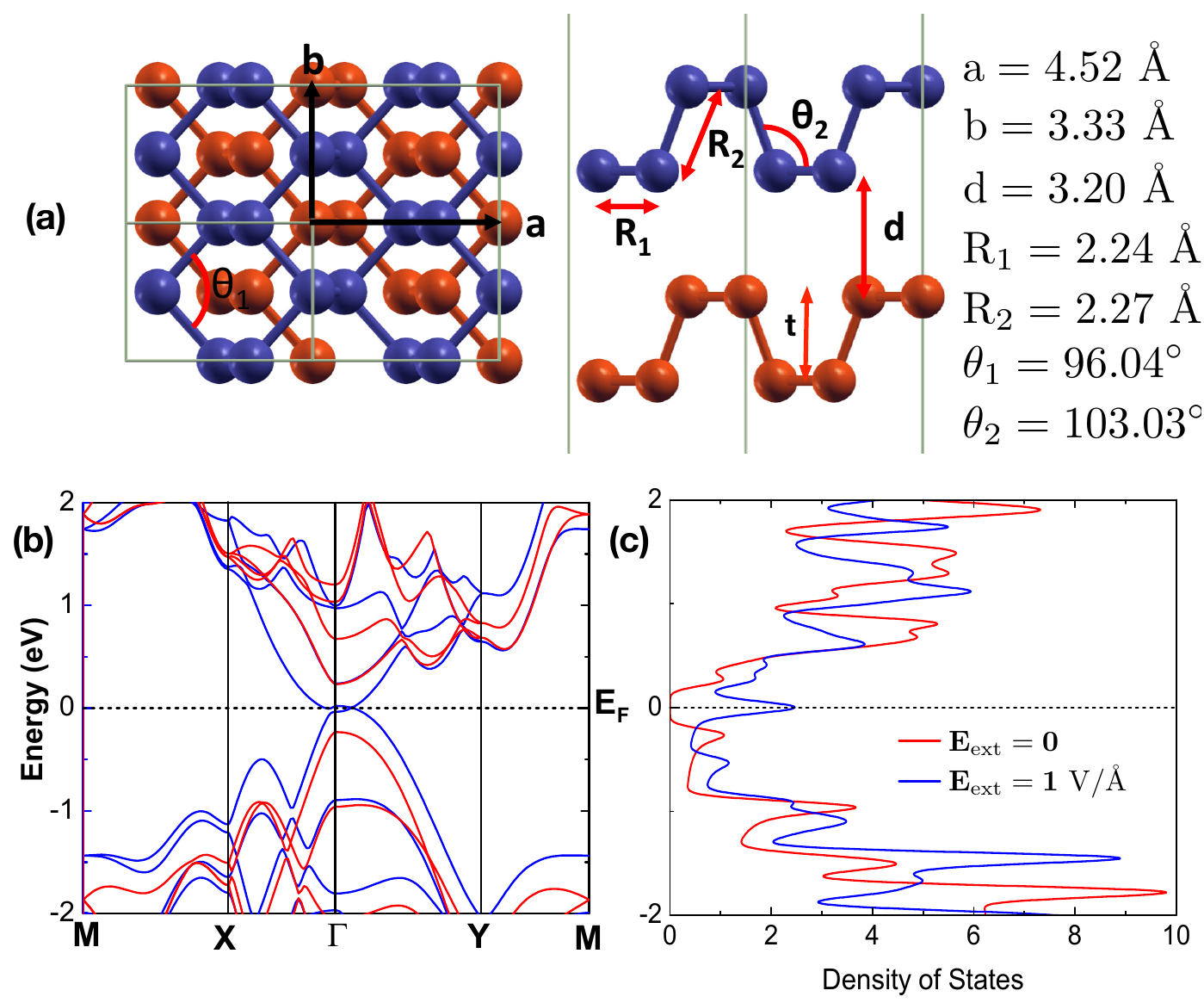}
\caption{(a) Top view and side view of bilayer phosphorene (BP) for AB stacking of layers, calculated using optB88-vdW. (b) Band structure and (c) Density of States of BP, in the absence (red curve) and presence (blue curve) of a transverse electric field of $1.0 $V/\AA. The red curve clearly shows an intrinsic direct bandgap of 0.48 eV at the $\Gamma$ point. Note that in presence of a large transverse electric field (blue curve), the bandgap reduces to zero signifying a semiconductor to metal transition which is also known to be a topological phase transition.\cite{PTI}  Structural properties for bilayer AB stacked phosphorene are consistent with those reported earlier.\cite{dai2014bilayer}}
\label{Fig.1}
\end{figure}

In Fig~\ref{Fig.1}(a), we illustrate the crystal structure and the electronic band structure of AB stacked bilayer phosphorene. Comparing with their bulk values, we find that in bilayer phosphorene the lattice parameter $a$ is 0.9\% larger, while $b$ and inter-layer distance $d$ remains the same [see Table~\ref{t1}]. With increasing number of layers, magnitude of $a$ decreases towards it's bulk value of 4.48~\AA~ [see Table~\ref{t1}]. The GGA-PBE estimated bandgap in bilayer phosphorene has a value of 0.48 eV, which decreases towards it's bulk value of 0.04 eV, the magnitude being inversely proportional to the number of layers [see Table~\ref{t1}]. Although GGA-PBE can qualitatively predict the electronic band structure of a material correctly, it's limitations are well known regarding bandgap underestimation. For example, a value of 1.04 eV (almost twice than our GGA-PBE estimate) is reported for bilayer phosphorene, based on hybrid functional (HSE06) based calculations, which is known to predict bandgap more accurately.\cite{dai2014bilayer} Evidently the band structure is anisotropic (see curvature  along $\Gamma$-X  and $\Gamma$-Y directions), which is also going to be reflected in effective mass, mobility etc. and this qualitatively explains the experimentally observed strong directionality of electronic, thermal and optical properties of multilayered black phosphorene.\cite{liu2014phosphorene,wang2015highly,mao2015}

\begin{table}[t]
\centering
\begin{tabular}{c c c c c}
\hline
NL & a (\AA) & b (\AA) & d (\AA) & Bandgap (eV)\\
\hline
2 & 4.52 & 3.33 & 3.21 & 0.48\\
3 & 4.51 & 3.33 & 3.21 & 0.28\\
4 & 4.51 & 3.33 & 3.21 & 0.19\\
5 & 4.50 & 3.33 & 3.21 & 0.13\\
6 & 4.50 & 3.33 & 3.21 & 0.09\\
8 & 4.49 & 3.33 & 3.21 & 0.07\\
Bulk & 4.48 & 3.33 & 3.21 &0.04\\
\hline
\end{tabular}
\caption{Structural parameters and bandgap of multi-layered and bulk black phosphorus calculated using GGA-PBE.}
\label{t1}
\end{table}

\section{Impact of a transverse electric field and charge screening} 
\begin{figure}[t!]
\includegraphics[width=0.8 \linewidth]{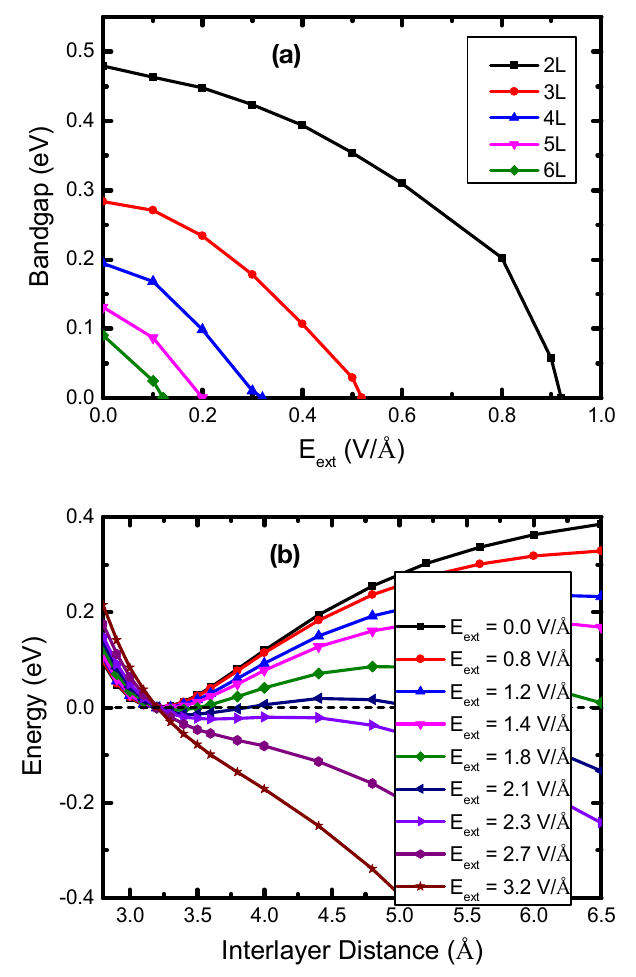}
\caption{(a) Variation of the (GGA calculated) bandgap with the applied electric field for multi-layered (two to six layers) phosphorene. (b) Total energy of bilayer phosphorene, calculated with reference to it's energy at equilibrium interlayer spacing of 3.2 \AA~ and zero external field, plotted as a function of interlayer distance for different $E_{\rm ext}$. Reduced stability of bilayer AB stacked phosphorene suggests that it can be more easily exfoliated at higher electric fields.}
\label{fig2}
\end{figure}

Having described the structural and electronic properties of multi-layered phosphorene, we now proceed to study the impact of a transverse electric field on it. As reported in the literature \cite{dai2014bilayer,dolui2015quantum, PTI}, with increasing external transverse electric field ($E_{\rm ext}$) in multilayered phosphorene, its  bandgap decreases and ultimately reduces to zero [see Fig.~\ref{fig2} (a)], leading to a topological phase transition with a phase characterized by the $Z_2=1$ index \cite{z2kane,z2Rahul2,z2Rahul,PhysRevB.83.235401}. However, in this article we will focus on the evolution of the dielectric properties of multi-layered phosphorene in normal insulator state, {\it i.e.}, before it turns into a topological insulator and eventually into a metal at relatively high transverse electric field.\cite{PTI} 
According to our GGA-PBE calculation, for bilayer phosphorene bandgap closing occurs at $E_{\rm ext} = 0.93 $V/\AA~[see Fig.~\ref{Fig.1}(b)-(c)]. The explicit dependence of the bandgap on the transverse electric field is shown in Fig.~\ref{fig2}(a). Evidently the critical electric field needed to close the bandgap and induce a topological phase transition decreases with increasing number of layers, which is consistent with the fact that the actual bandgap also varies inversely with number of layers. 

\begin{figure}[t!]
\centering
\includegraphics[width=0.9 \linewidth]{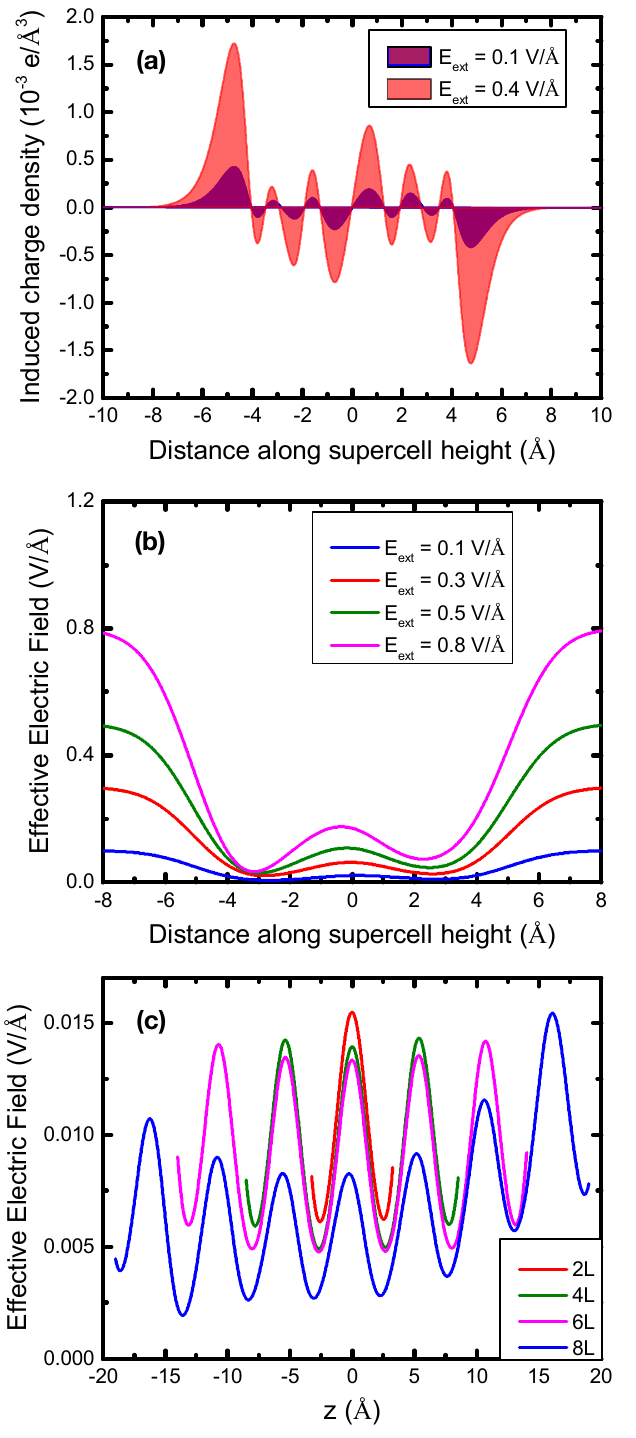}
\caption{(a) The induced charge density for bilayer phosphorene, within the supercell,  at two different $E_{\rm ext}$. The two extreme peaks become more asymmetric at higher electric field as depicted in the figure. (b) Variation of the effective electric field inside bilayer phosphorene at different $E_{\rm ext}$. (c) The effective electric field  inside few layer phosphorene as function of distance along $z$ axis for 2 to 8 layers at a fixed $E_{\rm ext}$ = 0.08 V/\AA.}
\label{fig3}	
\end{figure}

There is a limit to the electric field strength that can be applied to few layers of phosphorene without compromising its structural stability. Binding among adjacent layers [because of van-der Waals (vdW) interaction] are expected to weaken in presence of externally applied electric field, making the structure unstable beyond certain field strength. This is verified by computing the total energy of bilayer phosphorene (measured with reference to it's energy at equilibrium interlayer spacing of 3.2 \AA~ and zero external field) by changing the interlayer distance,  for different values of $E_{\rm ext}$ [see Fig.~\ref{fig2}(b)]. The vdW binding energy per phosphorus atom in absence of external electric field is about $48.1$ meV, while that at $E_{\rm ext} = 1.4$ V/\AA, it is $21$ meV and it further decreases to zero at $E_{\rm ext} = 1.8$ V/\AA.

Qualitatively, a vertical electric field pushes the electronic cloud from the bottom surface towards the top surface. This in turn results in induced charge density as depicted in Fig.~\ref{fig3}(a),  which is qualitatively similar to the induced charge density reported in other 2D materials such as graphene \cite{yu2008ab,santos2013electric}, MoS$_2$ \cite{santos2013electrically}, multilayer GaS films \cite{GaS} {etc.} To calculate the effective induced charge density in the $z$ direction (perpendicular to the phosphorene layers), denoted by $\rho_{\rm ind}({\bf r})$,  we take the difference between $\rho_{\rm ext}({\bf r})$ (charge density in presence of external field) and $\rho_0({\bf r})$ (charge density in absence of external field), and then average it over the plane of phosphorene layer ({\it x-y} plane). In order to smoothen out the variations in the charge density over the inter-atomic distances, we further use a Gaussian (filter) smoothing function, having the width of the order of inter-atomic distance and thus we have 
\be \label{eq:rho}
\langle \rho _{\rm ind}(z)\rangle = \frac{1}{S} \int_{x} \int_{y} \left[\rho _{\rm ext}({\bf r}) - \rho _0({\bf r})\right] dxdy
\ee 
where S denotes the area of the supercell in the {\it x-y} plane and the $\langle ..\rangle$ indicates the spatial in-plane average with the Gaussian filter in the $z$ direction. This induced `macroscopic' charge density typically increases with increasing electric field. 

The induced charge can then be used to calculate the planar averaged and Gaussian filtered effective polarization in the $z$ direction 
using,
\begin{equation} \label{eq:pdef}
\frac{\partial \langle p_{\rm ind}(z)\rangle}{\partial z}  = - \langle \rho_{\rm ind}(z) \rangle~,
\end{equation}
along with the boundary condition that the polarization vanishes in the region of vanishing charge density.

The induced charges 
in turn screens the external electric field by generating a screening electric field ($E_{\rho}$), and consequently the effective electric field inside phosphorene is reduced to $E_{\rm eff}(z) = E_{\rm ext} - E_{\rho}(z)$. The effective electric field in the $z$ direction 
can be calculated  by solving the Poisson equation for the screening potential which arises form the induced charges,  
\be \label{eq:Eeff}
\frac{\partial\langle E_{\rm \rho}(z)\rangle}{\partial z}  = - \frac{\langle \rho _{\rm ind} (z) \rangle}{\epsilon _0}~,
\ee
along with the boundary condition that the induced electric field vanishes far away from the region of induced charges. 

From an {\it ab-initio} perspective, $E_{\rm eff} (z)$ can also be directly calculated by taking the planar average of the difference between Hartree potential (obtained from DFT calculations) at finite electric field and that at zero electric field  and differentiating with respect to the vertical distance $\it z$ \cite{santos2013electric,santos2013electrically}: $E_{\rm eff}(z) = -\partial_z V_{H}(z)$, where $ V_{H}(z) \equiv V_{H}^{E}(z) - V_{H}^0(z)$. 
We have checked that the effective electric fields, calculated using both the methods are  consistent with each other (within $1 \%$ or each other), in the region where there are finite charges, and consequently we will just use  Eq.~\eqref{eq:Eeff} to report the effective electric field used in the rest of the manuscript. 

The variation of $E_{\rm eff}$ inside bilayer phosphorene, as function of position along {\it z}-axis at different electric field strengths is shown in panel (b) of Fig.~\ref{fig3}. Evidently both $E_{\rm eff}$, as well as the difference of $E_{\rm eff}$ felt by the two layers, increases with increasing applied electric field strength. Aa a consistency check we note that the $E_{\rm eff} \to E_{\rm ext}$, at the slab boundaries where the induced charges vanish. 
In panel (c) of Fig.~\ref{fig3}, we study the variation of  $E_{\rm eff}$ for different number of layers of phosphorene, as a function of the vertical distance, for a constant $E_{\rm ext} = 0.08$ V/\AA. Note that as the number of layers increase, the effective electric field decreases on account of the increased screening with thickness. We further note that with increasing number of layers,  the effective electric field is higher at the surface and decreases gradually  inside the material. This is a direct consequence of the fact that at the surface the field is screened only by the charges at the surface but as we move inside the material, the charges on the surface as well as those inside the material screen the external field. 

\begin{figure}[t!]
\includegraphics[width=0.99\linewidth]{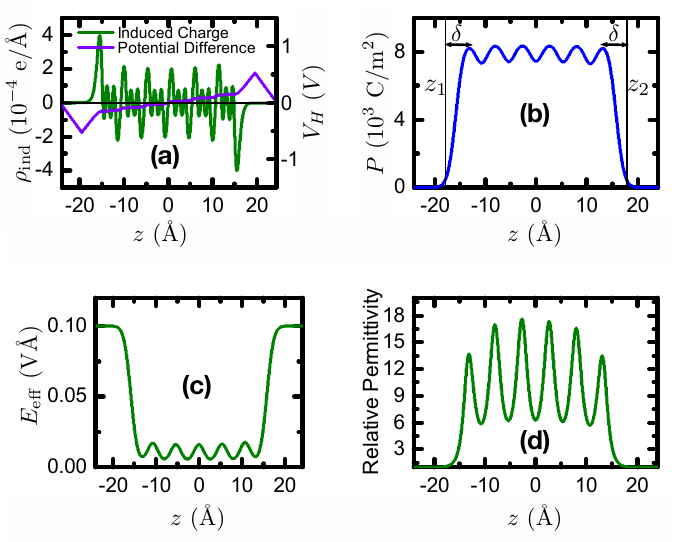}
\caption{(a) Induced charge density (without macroscopic smoothing out) and induced Hartree potential, $V_{H}(z) $, due to the external electric field, (b) polarization with the vertical lines marking the slab boundaries where the polarization falls to 1\% of the nearest peak values [see Eq.~\eqref{eq:er}], (c) effective electric field and (d) the relative permittivity, as a function of the distance along $z$ axis.  All panels are for six layer phosphorene for a transverse applied field of $E_{\rm ext} = 0.1$ V/\AA ~and  except for panel (a) all other panels show gaussian filtered quantities.}
\label{Dielectric_6L}
\end{figure}

\section{Polarizability and the dielectric function}

Having discussed the induced charge and the screened effective field, we now focus on the dielectric response of $N$-layer phosphorene. Following  Ref.~[\onlinecite{giustino2005theory}], the microscopic static permittivity (planer averaged with gaussian filter) can be defined as 
\begin{equation} \label{epdefn}
\epsilon _{r}(z) = 1 + \frac{\langle p_{\rm ind}(z) \rangle}{ \epsilon_0 \langle E_{\rm eff} (z)\rangle}~.
\end{equation}  
Using Eqs.~\eqref{eq:pdef}-\eqref{epdefn}, we can easily show that
\be \label{eq:continuity}
\partial_z \left[\epsilon(z) \langle E_{\rm eff}(z)\rangle \right]  = 0~, 
\ee
which is analogous to the conservation of the displacement field component perpendicular to the interfaces in electrostatics \cite{griffiths}. Equation~\eqref{eq:continuity} is also the basis for defining the inverse of the permittivity of a slab (of height $z_2 -z_1$) as the average of the inverse of the height dependent permittivity \cite{giustino2005theory}
\begin{equation} \label{eq:er}
\frac{1}{\epsilon_{r}} =  \frac{1}{z_2-z_1}\int_{z_1}^{z_2}\frac{1}{\epsilon _r(z)}dz ~.
\end{equation}
For this manuscript, we consider the slab thickness to be between the points where the polarization drops to $1\%$ of the nearest  peak value associated with the topmost or bottommost phosphorene layers, as shown in Fig.~\ref{Dielectric_6L}(b). 

\begin{figure}
	\includegraphics[width=0.9 \linewidth]{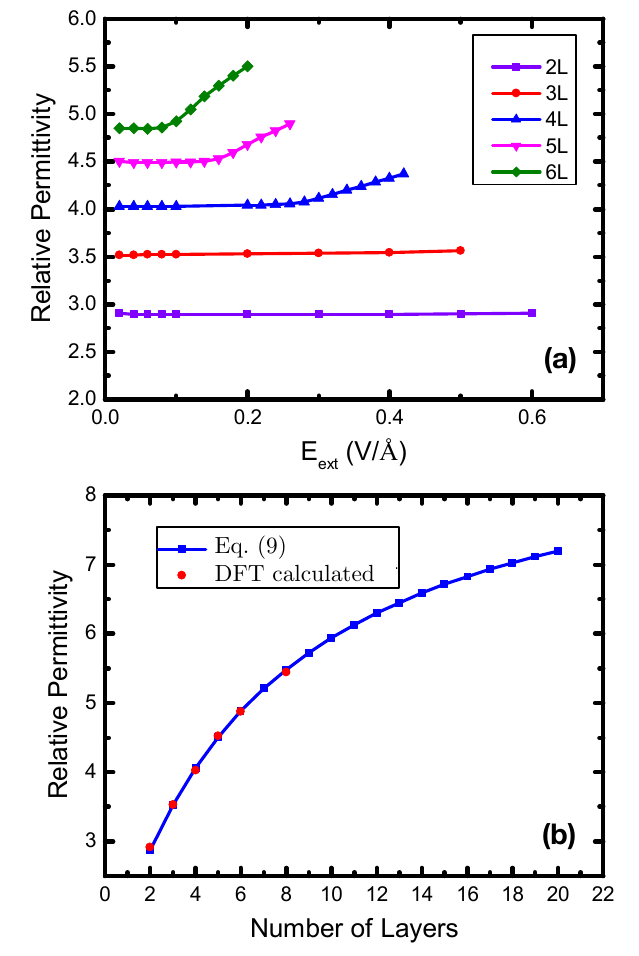}
	\caption{(a) Variation of relative permittivity with applied electric field for varying number of layers. (b) Variation of relative permittivity with number of layers at fixed external electric field ($E_{\rm ext} = 0.02 $V/\AA). The red circles represent the DFT calculated values, and the blue line represents the predicted values for any number of layers as per Eq.~\eqref{model}, which should yield good results for low electric fields where the electronic polarization is expected to vary linearly with the field,  and the ionic contribution to the effective field is negligible. 
\label{Permittivity_Eext}}
\end{figure}

Using the framework described above in Eqs.~\eqref{eq:rho}-\eqref{epdefn}, we calculate and display  the variation of the induced charge $\rho _{\rm ind}$ and the induced effective potential  ($V_H$) as a function of the vertical distance in six layer phosphorene for a fixed external field of $E_{\rm ext} = 0.1$ V/\AA~ in Fig.~\ref{Dielectric_6L}(a). In panels (b) and (c) of Fig.~\ref{Dielectric_6L}, we study the variation of the (Gaussian filtered) polarization and the effective induced field, respectively,  in the vertical direction. Panel (d) of Fig.~\ref{Dielectric_6L} shows the variation of relative permittivity $\epsilon _r(z)$ within the supercell in the vertical direction.

The dependence of the relative permittivity of multilayered black phosphorene,  defined by Eq.~\eqref{eq:er} as a function of the number of layers is displayed in Fig.~\ref{Permittivity_Eext}(a). Note that the relative permittivity for  bi- and tri-layer phosphorene is almost  constant and independent of the transverse electric field strength ($\epsilon_r = 2.9$, and $3.5$, respectively for bi- and tri-layer phosphorene) as long as phosphorene is in the insulating state. Further, even for four to six layered phosphorene, the relative permittivity is almost  constant for small electric fields and varies with the transverse electric field only at larger field strengths (roughly $E_{\rm ext} \ge 0.1$ V/\AA). In Fig.~\ref{Permittivity_Eext}(b) we study the variation of $\epsilon _r$ with number of layers for smaller electric fields. Evidently, the relative permittivity increases with the number of layers, and seems to be saturating slowly with increasing  number of layers towards its bulk value of $\epsilon_r = 8.3 $ for black phosphorus \cite{Asahina,Nagahama}.

\begin{figure}
	\includegraphics[width=0.91 \linewidth]{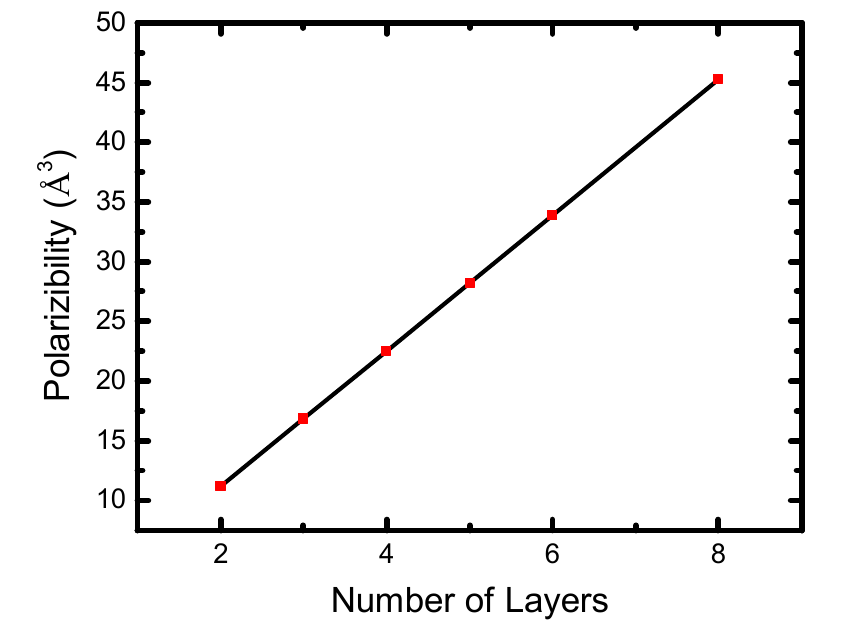}
	\caption{Slab polarizability as a function of the number of layers (red squared), which fits very well with a straight line (black). The linear behaviour of the slab polarizability with number of layers for low electric fields, implies that the polarizability per phosphorus atom is nearly constant and it has a value of $1.42 \times 4 \pi \epsilon_0$ \AA$^3$. For a comparison, the polarizability of a carbon atom in multilayered graphene sheet has a value\cite{yu2008ab} of $0.5 \times 4 \pi \epsilon_0$ \AA$^3$ . Here $E_{\rm ext} = 0.02 $V/\AA.
	\label{alpha_NL}}
\end{figure}

Besides the relative permittivity, atomic/molecular polarizability which relates the induced charge polarization or alternately dipole moment (in the $i$ direction) to the local effective electric field (in the $j$ direction) via the relation, $p_{{\rm ind},i} = \sum_j \alpha_{ij} E_{{\rm loc},j}$ is also of  great interest. Note that here, $E_{\rm loc}$ is the local electric field within the slab, which is in turn related to the macroscopic field $E_{\rm eff}$ in the slab, via the relation $E_{\rm loc} = E_{\rm eff} + p_{\rm ind}/{\epsilon_0} $, where $p_{\rm ind}$ is the average induced polarization in the slab.
The molecular polarizability $\alpha_{ij}$ is typically a tensor, however for this manuscript, we will focus on the polarizability in the $z$ direction for a transverse electric field in the $z$ direction, {\it i.e.}, $\alpha = \alpha_{zz}$. Now it is straightforward to relate the slab polarizability (in the $z$ direction) $\alpha$ of an isolated  phosphorene stack of $N$ layers, to the respective relative permittivity \cite{yu2008ab,li2015ab,tobik2004electric}, as 
\begin{equation} \label{eq:alpha}
\alpha(N) = \epsilon_0 \Omega_{\rm cell} \left(\frac{\epsilon _r(N) - 1}{\epsilon _r(N)} \right)~,
\end{equation} 
where $\Omega_{\rm cell}$ is the extended supercell volume to include the surface charges which extend beyond the top and bottom atomic phosphorene layers by $\delta$ [see Fig.~\ref{Dielectric_6L}(b), and Eq.~\eqref{Cell}]. In Eq.~\eqref{eq:alpha}, $\epsilon_r \to 1$ implies $\alpha \to 0$ which represents a slab of vacuum. Further in Eq.~\eqref{eq:alpha}, $\epsilon_r \to \infty$ implies $\alpha \to \epsilon_0 \Omega_{\rm cell}$, which is the polarizability of a metal slab. In general for an insulator,  finite polarizability is proportional to the fraction of the supercell volume that acts like a metal under a transverse electric field. Thus, if the polarizability per atom in phosphorene does not vary much, then the slab polarizability is expected to be a linear function of the number of layers. Note that Eq.~\eqref{eq:alpha} is valid only under the assumption that polarization is linearly proportional to the applied electric field, and further that there is no ionic contribution to the local electric field, both of which are likely to be valid only for small values of the applied electric field. We plot the polarizability as a function of the number of layers in Fig.~\ref{alpha_NL} and find that indeed the slab polarizability is a straight line, as expected.  
Similar linear behaviour of the slab polarizability with the number of layers,  has been demonstrated earlier for other two dimensional materials, such as benzene slabs \cite{tobik2004electric}, graphene \cite{yu2008ab}, GaS \cite{li2014modulation} etc. 

Following Refs.~[\onlinecite{tobik2004electric, yu2008ab, li2015ab}], the slab polarizability can also be expressed in terms of the bulk dielectric permittivity of black phosphorus $\epsilon _{\rm bulk}$ via the relation  
\begin{equation} \label{eq:alpha_b}
\alpha(N) = N \Omega_{\rm bulk} \frac{\epsilon_0}{2}\left( \frac{\epsilon_{\rm bulk} - 1}{\epsilon_{\rm bulk}} \right) + 8 \pi \epsilon_0 \alpha _s~,
\end{equation}
where $\Omega_{\rm bulk}$ is  the  volume of the bulk unit cell with two layers,  $\epsilon_{\rm bulk}$ is the dielectric constant of bulk black phosphorus,  and $\alpha _s$ is surface polarizability, which captures the difference in the dielectric properties of a slab and bulk. Based on the straight line fit to the slab polarizability data in Fig.~\ref{alpha_NL}, we obtain the value of  
$\epsilon _{\rm bulk} = 9.3$ and $\alpha _s = 0.09 $ \AA$^3$. Evidently, the surfaces are less polarizable than the bulk.
Here we emphasize that the bulk value of the relative permittivity of black phosphorene from the polarizability, is more than that of the actual value of 8.3 \cite{Asahina,Nagahama}. This discrepancy of less that $15\%$ in covalent solids, is known to be a direct consequence of simplistic assumption, of a solid being a collection of independently polarizable bi-layers which do not interact with each other, that is used to derive Eq.~\eqref{eq:alpha_b}. Similar discrepancy has also been  reported earlier for graphene \cite{yu2008ab} and Si \cite{giustino2005theory}. 

Finally we note that Eqs.~\eqref{eq:alpha}-\eqref{eq:alpha_b} can also be combined to predict the dielectric permittivity for any number of layers, once we have the results for a few layers, at least in the low electric field regime where the polarization is expected to be linearly dependent on the external field. Eliminating $\alpha(N)$ form Eqs.~\eqref{eq:alpha}-\eqref{eq:alpha_b}, we obtain 
\be \label{model}
\epsilon_r(N) = \left[1 - \frac{N}{2}\frac{\Omega_{\rm bulk}}{\Omega_{\rm cell}(N)} \frac{\epsilon_{\rm bulk} -1}{\epsilon_{\rm bulk}} - \frac{8 \pi \alpha_s}{\Omega_{\rm cell}(N)} \right]^{-1}~, 
\ee
where 
\be \label{Cell}
\Omega_{\rm cell} (N) = \Omega_{\rm bulk} \frac{N-1}{2} + \frac{\Omega_{\rm bulk}}{2(d+t)} (t + 2 \delta)~.
\ee
In Eq.~\eqref{Cell}, $t$ is the is the thickness of a single phosphorene layer due to its puckered nature and $d+t$ indicates the interlayer spacing [see Fig.~\ref{Fig.1}], and $\delta$ is the distance between $z_1$ (or $z_2$) and the exact supercell boundary {\it i.e.} where the bottommost (or topmost) atoms are located, as depicted in Fig.~\ref{Dielectric_6L}(b).  Physically $\delta$ indicates the region of charge spillover beyond the outermost atomic layers, and it does not vary significantly  with varying number of layers for low electric fields.
In Fig.~\ref{Permittivity_Eext}(b), we display the predicted effective dielectric constant from Eq.~\eqref{model} for up-to 20 layers, and evidently it shows a very good match with the actual DFT calculated values. 

\section{Conclusion}
To summarize, we have studied the dielectric properties of few layer black phosphorene, using first principle electronic structure  calculations and find that (a) in general the relative permittivity increases with increasing number of layers, ultimately saturating to the bulk value, and (b) it can be tuned to a certain extent by an external electric field.  The decreasing dielectric constant with reducing number of layers of phosphorene in the low field regime, is a direct consequence of the lower polarizability of the surface layers and the increasing surface to volume ratio.

In addition to the effective dielectric constant, we also calculate the slab polarizability of multilayered black phosphorene, and find that for small electric fields it displays a linear relationship with the number of layers, implying a nearly constant polarization per phosphorus atom
as expected.  However for large electric fields, which decreases with increasing number of layers, for example $E_{\rm ext} > 0.1$ V/\AA ~for 6 layers, and $E_{\rm ext} > 0.2$ V/\AA~ for 4 layers, as per GGA calculations --- see Fig.~\ref{Permittivity_Eext}(a), this simple relationship breaks down, on account of possible breakdown of the linear relationship between polarization and the applied field, and increasing impact of ionic contribution to the effective field. 
In the low field regime, one can model the system as a collection of well spaced and independent bi-layer units, and obtain the slab polarizability in terms of the bulk dielectric constant  and surface polarizability on one hand, and on the other the slab polarizability can be related to the layer dependent dielectric constant. This allows us to extrapolate our calculations for few layers, to much larger number of layers and obtain an empirical relation for the thickness dependent dielectric constant for any number of layers -- see Eq.~\eqref{model}.

Finally we note that all the calculations presented in this article, are valid only for the insulating regime of phosphorene. Further in a realistic experimental scenario, the effective dielectric constant is also likely to be affected by the environment, for example the substrate used, passivation method, etc. and incorporating these effects would require a more detailed numerical study. However our study offers a good starting point, and demonstrates the thickness dependent and electric field tunability of the dielectric constant in phosphorene, which provides an additional handle for optimally designing, and interpreting phosphorene based devices. 

\bibliography{ph_v2}
\end{document}